\documentclass[aps,pre,preprint,12pt,altaffilletter,showpacs]{revtex4}
\usepackage{graphicx}
\usepackage{amssymb}
\usepackage{amsmath} 
\newcommand{\sgn}{\mathop{\mathrm{sgn}}}
\usepackage{hyperref}
\usepackage{textcomp}
\usepackage{psfrag}
\usepackage{setspace}
\usepackage{url}
\begin{document}
\title{A Mean Field Model of Layering Instability in Shearing Suspensions}
\author{J.~I.~Katz$^*$}
\affiliation{Department of Physics and McDonnell Center for the Space
Sciences\\Washington University, St. Louis, Mo. 63130}
\email{katz@wuphys.wustl.edu}
\begin{abstract}
Concentrated suspensions may shear-thin when the suspended particles form
planar sheets that slide over one another with less friction than if the 
particles are randomly distributed.  In a na\"ive model the suspension is
described by a mean effective viscosity, and particles that collide with
each other redistribute the mean density in the shearing direction.  This
leads to a diffusion equation for the particle density.  If the viscosity in
the unthinned state is a steeply increasing function of particle density the
effective diffusion coefficient is negative and the diffusion equation,
meaningful only on scales larger than the particle separation, is ill-posed.
This singularity corresponds to the formation of planar sheets of particles
and defines a critical particle density for the onset of shear thinning.
\end{abstract}
\maketitle
Shear thinning of suspensions is ubiquitous.  It is, at least in part,
attributed to the formation of layers of suspended particles in planes
normal to the velocity gradient \cite{H72}, although the relation between
layering and shear thinning is complex \cite{SP05,CWJCY10,XRD13}.  Recent
theoretical work on this phenomenon \cite{BK11,KB11} has applied
thermodynamic methods (dynamic density functional theory) to colloidal
suspensions in which Brownian motion is significant.  At high P\'eclet
number a wall induces layering that propagates (at a rate not calculated)
into the bulk.  Yet shear thinning is also observed in non-Brownian
suspensions, such as those of corn starch \cite{H72,F08}, so it must also be
explicable by non-thermodynamic methods.

Shear-induced diffusion resulting from particle collisions can make
particles migrate through a rheometer and thereby decrease the measured
viscosity \cite{LA87}, an effect distinct from shear thinning.  Here I
present a mechanism by which particle collisions in sheared bulk suspensions
induce structure \cite{H72} on the scale of the particle size, and to shear
thinning, on time scales ${\cal O}(1/{\dot \gamma})$, where $\dot \gamma$ is
the mean (macroscopic) shear rate.  This is a much shorter time scale than
that of shear-induced transport \cite{LA87}.

Consider a simple model of shear thinning in a suspension of neutrally
buoyant particles interacting with only a hard-sphere repulsion.  Shear
thinning must be distinguished from thixotropy \cite{M79,B97} that results
from the breakup of clusters of mutually attracting particles, but these
do not exist in this hard-sphere model.  Unlike earlier models of sheared
colloids \cite{BK11,KB11}, the model does not assume the presence of a wall.
Shear stress may be a boundary condition at a wall, but here it is treated
as an initial condition in an unbounded suspension. 

Assume a steady plane creeping flow in the $\hat x$ direction, with velocity
gradient ${\dot \gamma} = {\partial v_x(z) \over \partial z}$ in the $\hat z$
direction and a constant shear stress $\sigma_{xz}$, and drop subscripts for
convenience.  Implicitly, $v$ and $\sigma$ are mean quantities averaged over
scales of the size of the particles, taken to be spheres of radius $a$ and
mean number density $n(z)$.  The suspending fluid has a viscosity $\eta_f$,
and the suspension is described as having a fluid viscosity $\eta$, again
averaging over its structure on scales ${\cal O}(a)$ in this mean field
model.

In a concentrated suspension multi-particle interactions are significant,
but they are not
calculable or even approximable analytically, so I only consider two-body
interactions.  The error introduced by this approximation is likely
comparable to that introduced by the approximate treatment of two-body
interactions.  The purpose of this treatment is necessarily only to provide
a qualitative explanation of the occurrence of layering and shear thinning,
and not a quantitative prediction of the properties of the layered state. 

Consider two spheres initially with centers $(x_1,z)$ and $(x_2,z + \Delta
z)$, $\vert \Delta z \vert < 2 a$ and $(x_1 - x_2) \Delta z {\partial v
\over \partial z} > 0$.  If the spheres were to continue to move at constant
$z$ and $\Delta z$ they would collide.  In fact, in this low Reynolds number
flow they will interact hydrodynamically (and with many neighboring
particles in a concentrated suspension).  When they approach closely their
separation will shrink exponentially under the influence of the hydrodynamic
stress $\sigma = \eta {\dot \gamma}$ with a time constant ${\cal O}
(\eta_f/\sigma) = {\cal O}[\eta_f/(\eta {\dot \gamma})] \ll 1/{\dot
\gamma}$, where $\eta$ is the macroscopic suspension viscosity and the thin
film of fluid between the spheres is described by the fluid viscosity
$\eta_f$.  Hence, despite their hydrodynamic interactions, their interaction
involves (with any finite, even atomic-scale, surface roughness) a hard
sphere collision, at least for small enough $\vert \Delta z \vert$ (for
larger $\vert \Delta z \vert$ hydrodynamic interaction may prevent such a
close approach, a possibility we ignore).  The hard-sphere repulsion breaks
the kinematic reversibility of creeping flow \cite{LA87,P77} that would
otherwise ensure that after interaction the spheres return to the initial
values of their $z$ coordinates.

\begin{figure}
\begin{center}
\includegraphics[width=4.5in]{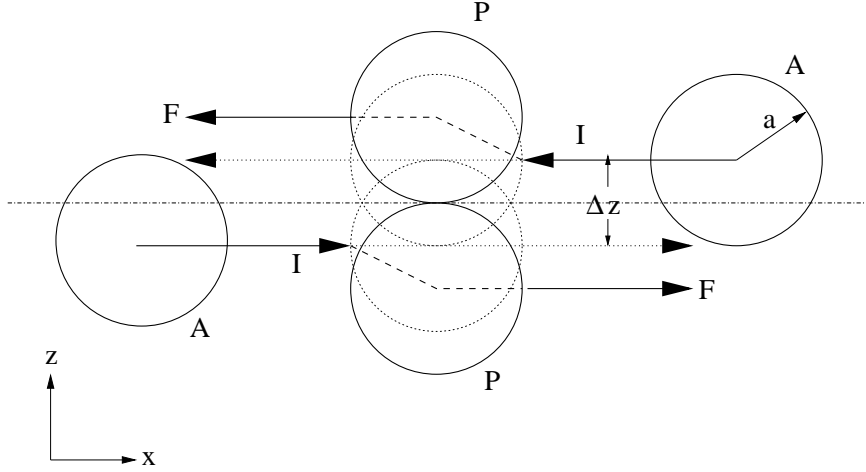}
\end{center}
\caption{Toy model of interaction of two spheres of radius $a$ in $y = 0$
plane in a shear flow $v_x = {\dot \gamma} z$.  The centers of the spheres
are initially on trajectories I separated by $\Delta z < 2a$.  A indicates
the locations of the spheres before interaction, F indicates the paths of
their centers after interaction, and P indicates their nominal positions at
closest approach.  The dashed lines indicate (but do not represent
quantitatively) their paths during interaction.  Dotted circles and lines
indicate overlapping positions and paths of undeflected (interpenetrating)
spheres.  The model is only qualitative because it neglects the complex
hydrodynamic interaction of the spheres with each other and with other
particles in a concentrated suspension.}
\label{collision}
\end{figure}

In a two-dimensional approximation, in which the spheres are confined to the
$y = 0$ plane (an approximation that simplifies the geometry of their
interaction without changing its qualitative properties), the consequence of
this collision is that when $x_1 = x_2$ the spheres touch ($\vert z_2 - z_1
\vert = 2a$) in order that they may pass each other; depending on the initial
$\Delta z$ they may come into contact even before then.  Again ignoring their
hydrodynamic interaction by assuming that $z_1$ and $z_2$ do not change as
the spheres move apart in the shear flow, the net displacements are 
\begin{equation}
\delta z = {\Delta z \over 2} - \sgn(\Delta z)a.
\label{deltaz}
\end{equation}
This geometry is shown in Fig.~\ref{collision}.

The rate of collisions
of a sphere with center $(x,z)$, taking all variables independent of $x$, is
\begin{equation}
\int_{\Delta z = - 2a}^{2a}\,d\Delta z \left\vert {\partial v \over \partial
z}\Delta z \right\vert n(z + \Delta z).
\end{equation}
The mean vertical velocity of this sphere is
\begin{equation}
\label{vz1}
v_z = \int_{\Delta z = - 2a}^{2a}\,d\Delta z \left\vert {\partial v \over
\partial z}\Delta z \right\vert n(z + \Delta z) \delta z,
\end{equation}
similar to a result of \cite{ZBW12}.  By convention, we take $\partial v /
\partial z > 0$.

In order to evaluate this expression we note that the assumption of constant
and uniform $\sigma$ (as found in steady planar flow) relates the velocity
gradient at $z + \Delta z$ to the viscosity there
\begin{equation}
{\partial v(z + \Delta z) \over \partial z} \approx {\sigma \over \eta_0 +
\left.{d\eta \over dn} \right\vert_{n=n_0} {\partial n \over \partial
z} \Delta z} \approx {\sigma \over \eta_0} \left(1 - \eta^\prime {\partial n
\over \partial z} {\Delta z \over n_0}\right),
\label{pvpz}
\end{equation}
where $\eta^\prime \equiv \left. {d \ln{\eta} \over d \ln{n}} 
\right\vert_{n=n_0}$, $n_0$ is the mean density of spheres per unit
area in the $y = 0$ plane, the viscosity has been expanded around $\eta(n_0)
\equiv \eta_0$ and $\Delta z$ and variables have been expanded to first
order around their mean values.  Integrating Eq.~\ref{vz1}, using 
Eqs.~\ref{deltaz}, \ref{pvpz}, and $n(z + \Delta z) \approx n_0 + {\partial
n \over \partial z} \Delta z$, and taking only the lowest non-vanishing
(linear) order in $\partial n \over \partial z$, we find
\begin{equation}
\label{vz2}
\begin{split}
v_z &= \int_0^{2a}\,d\Delta z {\sigma \over \eta_0} \left[\Delta z n_0
\left({\Delta z \over 2}-a\right) + {\partial n \over \partial z} \left(1-
\eta^\prime\right)\Delta z^2 \left({\Delta z \over 2}-a\right) + \ldots
\right]\\
&\quad -\int_{-2a}^0\,d\Delta z {\sigma \over \eta_0} \left[ \Delta z n_0
\left({\Delta z \over 2}+a\right) + {\partial n \over \partial z} \left(1-
\eta^\prime\right)\Delta z^2 \left({\Delta z \over 2}+a\right) + \ldots
\right]\\
&\approx {4 a^4 \sigma \over 3 \eta_0} \left(\eta^\prime - 1\right)
{\partial n \over \partial z},
\end{split}
\end{equation}
where the minus sign in the second term comes from the absolute value in
Eq.~\ref{vz1}.  This result is comparable to the equally approximate results
of \cite{BK11,KB11}.

The flow of particles satisfies the one-dimensional continuity equation
\begin{equation}
{\partial n \over \partial t} + n_0 {\partial v_z \over \partial z} = 0,
\label{continuity}
\end{equation}
where we have approximated $n \approx n_0$ because $v_z$ is first order in
small quantities and we are interested in infinitesimal perturbations from
a homogeneous state.  Substituting Eq.~\ref{vz2} into Eq.~\ref{continuity}
yields a diffusion equation
\begin{equation}
\label{diffeq}
{\partial n \over \partial t} + {4 a^4 \sigma n_0 \over 3 \eta_0}
\left(\eta^\prime - 1\right) {\partial^2 n \over \partial z^2} = 0.
\end{equation}
The diffusion coefficient resulting from collisions between particles
\begin{equation}
D_{zz} = {4 n_0 \sigma a^4 \over 3 \eta_0} \left(1 - \eta^\prime\right) =
{4 a^2 \phi \sigma \over 3 \pi \eta_0} \left(1 - \eta^\prime\right) = \phi
{\dot \gamma} {4 a^2 \over 3 \pi} \left(1 - \eta^\prime\right),
\label{deq}
\end{equation}
where the (two-dimensional) filling factor in the $y = 0$ plane $\phi = \pi
a^2 n_0$ and the stress $\sigma = \eta_0 {\dot \gamma}$.  To order of
magnitude, $D_{zz} \sim a^2 {\dot \gamma}$, as must be the case because this
is the only quantity with dimensions of diffusivity that can be formed from
the parameters of the problem.

In general, in a concentrated suspension the viscosity is a steeply
increasing function of $\phi$ \cite{SP05} so that we expect $\eta^\prime >
1$ and $D_{zz} < 0$.  The diffusion equation is ill-posed unless the
negative $D_{zz}$ is offset by a positive diffusivity.  At finite
temperature Brownian diffusion adds a positive $D_{th} \sim (kT/\eta_0 a)$,
where in a concentrated suspension the coefficient depends on the
environment.  However, the total diffusivity $D = D_{th} + D_{zz}$ will
still be negative (if $\eta^\prime > 1$) for $a$ greater than some threshold
corresponding to large P\'eclet number, and large enough $\eta_0(\phi)$,
corresponding to a sufficiently concentrated suspension.  The condition
$D < 0$ becomes
\begin{equation}
\label{criterion}
\eta_0(\phi) \gtrsim {3 \pi \over 4} {kT \over a^3 \phi {\dot \gamma}
(\eta^\prime - 1)}.
\end{equation}

If the inequality (\ref{criterion}) is satisfied the mathematical
catastrophe of ill-posedness is avoided because Eq.~\ref{deq} only describes
mean field fluid quantities such as $n_0$ and $\eta$ that are not defined on
scales $\lesssim a \phi^{-1/2}$.  The fastest growing perturbations are
those with wavelengths $\sim a \phi^{-1/2}$ that have $e$-folding times
\begin{equation}
t_{growth} \sim {a^2 \over D_{zz}\phi} \sim {1 \over \phi {\dot \gamma}
(\eta^\prime - 1)}.
\end{equation}

This instability will saturate at finite amplitude at which Eq.~\ref{diffeq}
breaks down because higher-order or nonlinear terms become significant.
The result is the formation of structure in the
$\hat z$ direction such as the sheets in $\hat x$-$\hat y$ planes observed
for shear thinning suspensions \cite{H72}.  The inequality (\ref{criterion})
is then the criterion for the onset of layering.  If the suspension is
non-Brownian (the limit $kT/a^3 \to 0$) then the instability criterion
reduces to $D_{zz} < 0$ or $\eta^\prime(\phi) > 1$, which is an implicit
criterion for $\phi$.  In contrast to the results of \cite{BK11,KB11,ZBW12},
this layering instability is a bulk phenomenon, and is not dependent on
confinement or hydrodynamic interaction with walls.

Allowing for separation of interacting particles in the third (vorticity)
dimension $\hat y$ would reduce $v_z$ and $D_{zz}$ by factors ${\cal O}(1)$
but would not change the conclusion that a layering instability occurs for
$\eta^\prime > 1$.  However, because there is no momentum flow in the $\hat
y$ direction ($\sigma_{xy} = 0$) there is no equation for $\partial v \over
\partial y$ analogous to Eq.~\ref{pvpz} and no instability producing
structure in that direction.
\begin{acknowledgments}
I thank H.~A.~Stone for discussions.  This work was supported in part by
American Chemical Society Petroleum Research Fund \#51987-ND9.
\end{acknowledgments}

\end{document}